\documentclass[epj,referee]{svjour}

\usepackage{graphicx}
\usepackage{fancyhdr}
\usepackage{subfigure}
\usepackage{indentfirst}

\usepackage{cite}
\usepackage{amssymb}
\usepackage{newlfont}

\begin{document}

\title{Identifying overlapping communities in social networks using multi-scale local information expansion}

\author{ Hui-Jia Li\inst{1,2} \and Jun-Hua Zhang\inst{1,2,3}\and Zhi-Ping Liu\inst{2,4} \and Luonan Chen\inst{2,4,5,}
\thanks{\emph{Email:} lnchen@sibs.ac.cn} \and Xiang-Sun Zhang\inst{1,2,}
\thanks{\emph{Email:} zxs@amss.ac.cn}%
}                     
%
%
\institute{ Academy of Mathematics and Systems Science, Chinese
Academy of Sciences, Beijing 100190, China. \and  National Center
for Mathematics and Interdisciplinary Sciences, Chinese Academy of
Sciences, Beijing 100190, China. \and Key Laboratory of Random
Complex Structures and Data Science, Chinese Academy of Sciences,
Beijing 100190, China. \and Key Laboratory of Systems Biology,
Shanghai Institutes for Biological Sciences, Chinese Academy of
Sciences, Shanghai 200233, China. \and Collaborative Research Center
for Innovative Mathematical Modelling, Institute of Industrial
Science, University of Tokyo, Tokyo 153-8505, Japan.}
\date{Received: date / Revised version: date}

\abstract{Most existing approaches for community detection require
complete information of the graph in a specific scale, which is
impractical for many social networks. We propose a novel algorithm
that does not embrace the universal approach but instead of trying
to focus on local social ties and modeling multi-scales of social
interactions occurring in those networks. Our method for the first
time optimizes the topological entropy of a network and uncovers communities
through a novel dynamic system converging to a
local minimum by simply updating the membership vector with very low
computational complexity. It naturally supports overlapping
communities through associating each node with a membership vector
which describes node's involvement in each community. This way, in
addition to uncover overlapping communities, we can also describe
different multi-scale partitions by tuning the characteristic size
of modules from the optimal partition. Because of the high
efficiency and accuracy of the algorithm, it is feasible to be used
for the accurate detection of community
structure in real networks.}

\maketitle

\section{ Introduction }

Since the publication of the seminal works of Barab\'{a}si and
Albert \cite{Barabasi}, a lot of real complex systems have been
examined from the viewpoint of complex networks. Having been
observed to arise naturally in a vast range of physical phenomena,
complex networks can describe complex systems containing massive
units (or subsystems) with nodes representing the component units
and edges standing for the interactions among them. The social
network is a representative complex network and closely related to
our life, for example, World Wide Web \cite{Albert}, traffic
networks \cite{Li}, sexual networks \cite{Liljeros}, and article
cite networks \cite{Sumiyoshi}.

The study on the community structure of social networks has become a
very important issue in the field of complex networks. Nodes, which
belong to a tight-knit community, are more likely to have particular
properties in common. It is significantly important to identify
communities in social networks. By taking $WWW$ network as an
example, groups of web pages are more likely linking to web pages on
related topics. These sets of web pages might correspond to some
kinds of communities. Based on this search engines may increase the
precision and recall of search results by focusing on narrow but
topically-related subsets of the web.

The problem of finding communities in social complex networks has
been studied for decades. Recently, several quality functions for
community structure have been proposed to solve this
problem\cite{Newman}\cite{Newman01}\cite{Newman02}. Among them,
modularity $Q$ is proved to be the most popular
\cite{Newman02}\cite{Danon}\cite{Our} and has been pursued by many
researchers \cite{Clauset}\cite{Newman03}\cite{Our2}. However, most of those
approaches require knowledge of the entire graph structure to
identify global communities based on global information. That means,
one needs to access the whole network information. This constraint
is impractical for large complex networks, because it is a challenge
to know the whole network completely. Moreover, statistical methods
can only detect the most significant connectivity community patterns
and ignore their multi-scale topology. These identifications don't
have the advantage of providing a coarse-grained representation in
the system, thereby they can't sketch its organization or identify
the sets of nodes which are likely to have hidden functions or
properties in common.

Because of these limitations, we present a novel algorithm for
community detection focused on social networks in this paper. The
algorithm does not embrace the universal approach instead of trying
to focus on social networks using local information and modeling the
multi-scale social interaction patterns occurring in those networks.
Our method optimizes the topological entropy that represents the
statistic significance of a network for the first time. Although the
topological entropy function is not convex and it is unrealistic to
expect a standard optimization algorithm to find the global minimum,
we develop a novel dynamic system which converges to a local minimum
by simply updating the membership vector with low computational
complexity. We don't need to specify the number of communities that need to partition.
It naturally supports overlapping communities by
associating each node with a membership vector describing node's
involvement in each community. Theoretical analysis and experiments
show that the algorithm can uncover communities fast and accurately.

The outline of this paper is as follows. Section 2 introduces the
problem of community detection in social networks and the motivation
behind our algorithm. In section 3, we present our algorithm through
four steps and explain each one respectively. In section 4, we
analyze some important properties of our algorithm and in section 5,
we run this algorithm in several real social networks which
represent a good fraction of the wide body of social networks.
Section 6 concludes this paper.

\section{ Motivation }

Given a network $G=(V,E)$ contains $n$ nodes, suppose we can divide
them into $a$ groups. For each group suppose that we can select a
``leader". The leaders should be able to have two properties: they
should be well connected to the members of their group, and they
should also be able to communicate with other leaders when
necessary. If the distributed algorithm is carried out in each group
separately and the leaders communicate on a higher level, the agents
can enjoy faster convergence rate.

\begin{figure}
\resizebox{1\columnwidth}{!}{\includegraphics{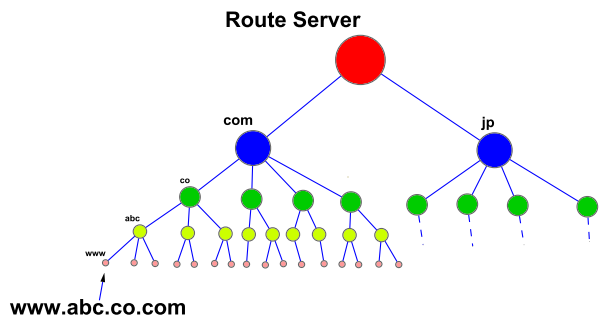} }
\caption{Hierarchical structure of DNS network with IP
``www.abc.co.com". The most influential node, router server, is
located on a highest level in the hierarchical tree. The servers
that include ``www", ``abc", ``co" and ``com" are located at lower
level. To obtain the IP address, users need to make a inquiry from
the highest level router server to lowest www server. Node size
depicts different levels in the hierarchy with the bigger node
locating at the higher level.}
\label{Fig:1}       
\end{figure}

It is natural to relate social networks with hierarchical structure.
In one such hierarchy there are leader nodes that are more important
than some other nodes, hence located on a higher level in the
hierarchy. It naturally follows that the leader is located on the
highest level within that hierarchy. By taking the DNS network in
$WWW$ \cite{Albert} as an example, the Router server is a natural
leader and locates on the highest hierarchy(see Fig.1) when
searching IP address. Since the hierarchies are consequence of the
spreading of its correlation, and so are the communities, we believe
that the identification of these hierarchies in a network will
result in a natural community detection. The area on which a leader
has most influence should define its community. So, community
detection is performed by finding all natural leaders and all nodes
on which they have influence. Partitions obtained in this way can be
naturally explained. Also, another intuitive property that a
community should possess is satisfied this way, that is shortest
paths exist between nodes from a same community.

Given a graph, individual nodes only have local knowledge about its
structure, which include information about their neighboring nodes.
If any node wants to improve its own performance, it needs to know
more about the global picture of the network. This information can
be used by the node to refine its choice of neighbors in order to
improve its performance. However, this will cost a lot of
computational complexity. The most complete measure of global graph
structure is the adjacency matrix. Since each node has limited
memory, energy, and computational capacity, it will be difficult to
directly use the adjacency matrix. Our goal is to devise a scheme to
provide each node with a small vector that includes compact global
information on how the node is located with respect to the other
nodes. It is desired that the scheme can be disseminated via an
implementable distributed manner.

Moreover, a powerful method uncovering the modules in social
networks should use a multi-scale way \cite{Lambiotte}\cite{Mucha}.
This identification has the advantage of providing a coarse-grained
representation of the system, thereby allowing to sketch its
organization and to identify sets of nodes that are likely to have
hidden functions or properties in common. Most community detection
methods find a partition of the nodes into communities, where most
of the links are concentrated within the communities. Each node is
assigned to one and only one community, i.e., partitions are not
compatible with overlapping communities \cite{Palla}\cite{Our1}. At
the heart of most partitioning methods, there is a mathematical
definition for what is thought to be a good partition. Once this
quality function has been defined, different types of heuristics can
be used in order to find, approximatively, its optimal partition,
i.e., to find the partition having the highest value of the quality
function.

\section{ The algorithm }

For a network $G=(V,E)$ with $n$ nodes, we develop a distributed
algorithm, which can categorize the node as ``leader" or ``regular"
using local information. Further, the method assigns each regular
agent with a membership vector in multi-scale way, indicates that
leaders has more influence on it. This provides the nodes with some
global picture of the network. The iteration includes three steps
described as follows.

\subsection{ Leadership of nodes }

First, we calculate the leadership $f_i$ of every node $i$ in the
network. The leadership $f_{i}$ represents how important is the
opinion of node $i$ in the network. Let the node leadership function
defined as:

\begin{equation} \label{eq:1}
f(i)=
\sum^{n}_{j=1;d_{ij}\leq\lfloor\frac{3\delta}{\sqrt{2}}\rfloor}e^{-\frac{d_{ij}}{\delta}}
\end{equation}
where $d_{ij}$ is the shortest distance from vertex $i$ to vertex
$j$. $\delta\in(0,+\infty)$ is the influence factor which is used to
control mutual action range between nodes. According to the
properties of exponential function $e^{-\frac{d_{ij}}{\delta}}$, for
a special value of $\delta$, the influence range of every node to
other nodes is approximately
$\lfloor\frac{3\delta}{\sqrt{2}}\rfloor$. When $d_{ij}$ larger than
$\lfloor\frac{3\delta}{\sqrt{2}}\rfloor$, the value of exponential
function rapidly reduce to $0$, so we can use $\delta$ to control
the influence range of a node and calculate $f(i)$ only within the
range $d_{ij}\leq\lfloor\frac{3\delta}{\sqrt{2}}\rfloor$. For the
dense region of a network, nodes have higher leadership. The nodes
with largest leadership mean they have most amount of links with
other nodes and can be viewed as candidate of leader nodes.
Therefore, we can use node leadership to represent the importance of
a node in the network.

\subsection{ Identifying the leader nodes }

Identifying the leader nodes of the community is very important to
analyze the properties of the complex networks. Many ways can be
used to define the ``key node", such as the nodes with largest
degree or betweenness centrality. Here, we use node leadership to
search leader nodes. According to the notion of community structure,
the density of inner-community links is larger than the rest of
nodes. Each community represents a local region with relative higher
correlation and the leader node of the community has the highest
leadership and is tightly linked by other nodes. Moreover, different
communities are divided by local lowest leadership nodes -- the
boundary nodes.

Note that in the rare cases where two or more leaders are also most
influential neighbors between each other, then they are grouping
together and are becoming leaders of one group. For example, in a
full connected network, all of the nodes are leaders of one
community, whereas for a ring network, each node is a leader to its
own community. Specifically, if the length of two highest leadership
nodes less than $\lfloor\frac{3\delta}{\sqrt{2}}\rfloor$, we group
them together and consider they are in one group. Finding leader
nodes only needs a simple breadth first search and if find, we
choose a random node to restart this process until converge. The
computational complexity is $O(m)$, where $m$ is the number of edges
in the network.

\subsection{ Determining the membership using random walk }

At this step, our goal is to devise a scheme to provide each node
with a small vector that includes compact global information on how
the node is located with respect to the other nodes. We provide a
definition for the membership vector based on the properties of
random walk dynamic on graphs. Consider a graph with $a$ leaders
${l_1,l_2,...,l_a}$ and $n-a$ regular nodes. Given the leaders and
the arbitrary order assigned to them, we describe the algorithm to
determine the membership vectors for each regular node. We denote
the membership vector of node $i$ by ${\textbf{x}_i}=(x_i^1, x_i^2,
\cdots, x_i^a)\in{R^a}$. By $x_i^k(t)$, we mean the $k$-th entry of
the influence vector of node $i$ evaluated at time $t$.

The procedure operates as follows. The membership vector of leader
$l_i$ is first assigned to be the unit vector. These $a$ vectors do
not vary. For regular node $i$, $x_i^k$ is initialized randomly,
distributed uniformly on $[0,1] (k=1,2,...,a)$. Then we normalize
each row of $\textbf{x}_i$ so that for all leader $k$, the sum of
$x_i^k$ is 1. At each iteration time $t$, the influence vector of
each regular node $i$ is updated entry-wise $(k=1,2,...,a)$ using
the following rule:

\begin{equation} \label{eq:2}
x_i^k(t+1)=\frac{1}{\sum_{j}a_{ij}+1}[x_i^k(t)+\sum_{j}a_{ij}x_j^k(t)]
\end{equation}
where $A=\{a_{ij}\}$ is the adjacency matrix in which $a_{ij}=1$ if
node $i$ and $j$ are connected and $a_{ij}=0$ otherwise.

We notice that, for all time $t$, $\sum_k x_i^k(t)=1$. Equation
$(2)$ is equivalents to $X(t+1)=PX(t)=(I + D)^{-1}(A + D)X(t)$,
where $P=(I + D)^{-1}(A + D)$ is a stochastic walk matrix. Actually,
the influence of leader nodes $l_k(k=1,2,...,a)$ on any regular node
$i$, $x_i^k$, is the probability that a random walker that starts
from $i$ hits $l_k$ before it hits any other leader
node\cite{Baras}. If the underlying graph is connected, the
iteration $\lim_{t\rightarrow{\infty}}x_i(t)$ converges to a set of
unique vectors and these vectors can naturally be represented as the
probability a regular node belongs to the community that a given
leader node in. As a result, although leadership of a node only
contain local information, we can use random walk dynamic to gain
membership containing a global view of the whole graph.

\section{ Some descriptions of the algorithm }

In this section we describe several important properties of the
algorithm, including computing the influence factor $\delta$ to
recognize multi-scale communities, identifying the leaders using
local information, determining the overlapping nodes and estimating
the complexity of the algorithm.

\subsection{ Determining the influence factor $\delta$ to recognize multi-scale communities }

According to the definition of leadership, the algorithm is
controlled by only one parameter, the influence factor $\delta$. We
can naturally use $\delta$ to control the scale of community
structure detected by our method. Here, we introduce topological
entropy $H$ \cite{Gfeller}\cite{Bianconi} that represents the
statistic significance of a network to choose suitable $\delta$: for
network $G=(V,E)$, $V={v_1,v_2,...,v_n}$, the leadership of $V$ are
$f(1),f(2),...,f(n)$, the topological entropy is defined as:

\begin{equation} \label{eq:3}
H=-\sum_{i=1}^{n}\frac{f(i)}{\sum_{i=1}^{n}f(i)}\log[\frac{f(i)}{\sum_{i=1}^{n}f(i)}]
\end{equation}

\begin{figure}
\centering
\subfigure[]{
    \label{fig:subfig:2a} 
    \setcounter{subfigure}{1}
    \includegraphics[width=7.5cm,height=5cm]{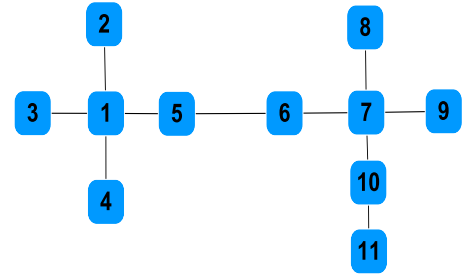}}
  \subfigure[]{
    \label{fig:subfig:2b} 
    \setcounter{subfigure}{2}
    \includegraphics[width=7.5cm,height=5cm]{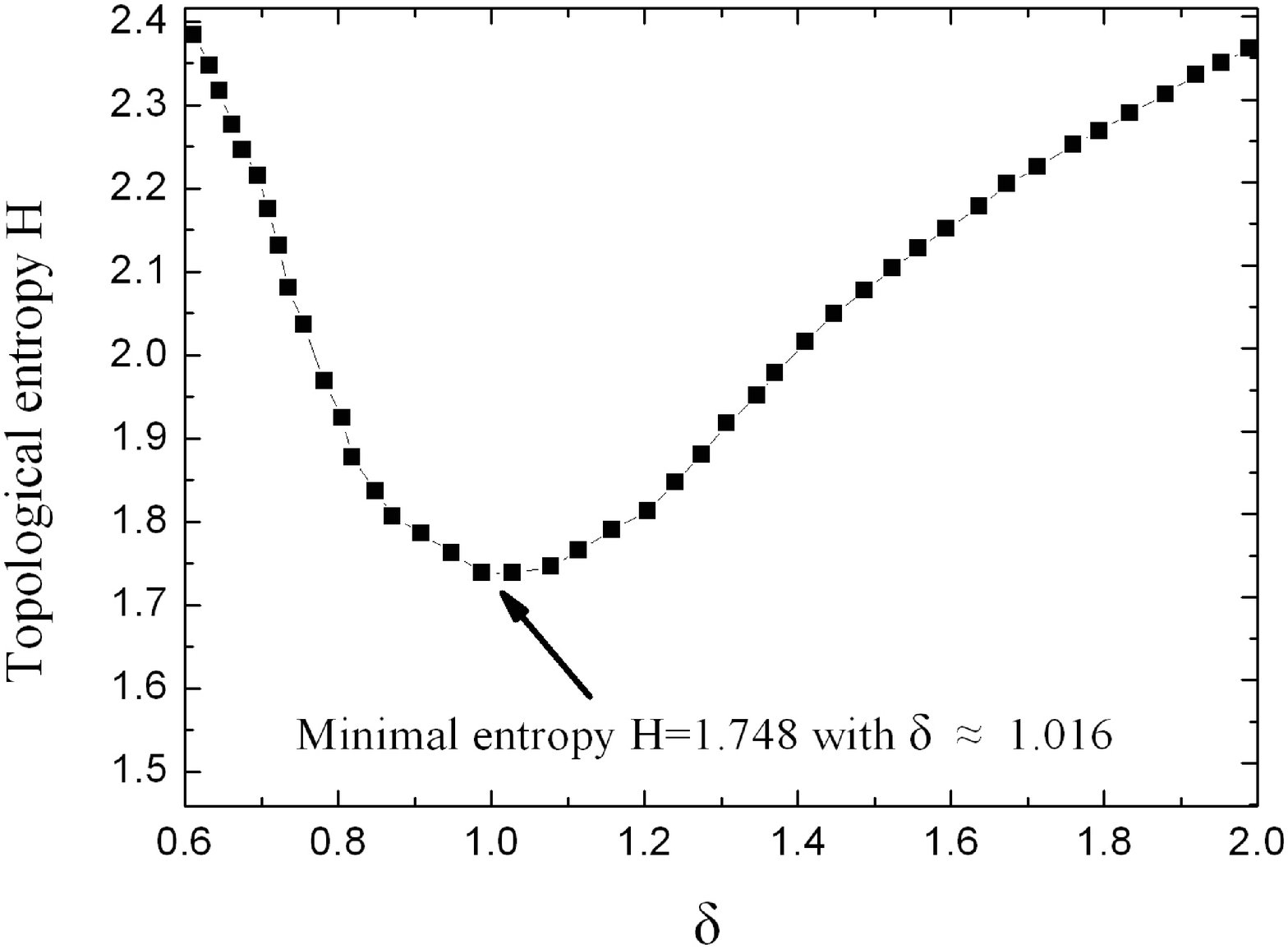}}
\caption{(a). A simple network with eleven nodes. (b). Plot of topological entropy $H$ versus influence factor $\delta$.}
\label{Fig:2}
\end{figure}

Small $H$ means a stable and suitable partition. For a simple
example, we consider network contain 11 nodes shown in
Fig.\ref{fig:subfig:2a} and calculate topological entropy
corresponding to different $\delta$. As shown in
Fig.\ref{fig:subfig:2b}, when $\delta$ increases from 0, the
corresponding entropy begins to decrease and reach minimal 2.2805 at
s specific value of $\delta$ $(\delta=1.26)$. When $\delta$ leaves
from optimal value, entropy begin to increase with $\delta$ and
finally reach the maximal value.

Therefore, to find a optimal $\delta$, it is equivalent to minimize
the single parameter nonlinear function $H(\delta)$ and many
algorithms can be used, for example, random search algorithm and
simulated annealing algorithm. However, $\delta$ corresponding to a
small value of $H$ but not minimal is also meaningful. Specially,
according to the property of leadership, the influence range of a
node is approximately $\lfloor\frac{3\delta}{\sqrt{2}}\rfloor$. When
$0<\delta<\sqrt{2}/3$, there is no interaction between two nodes.
Because no interaction exists, every node belongs to community
contains itself and the number of community is $n$.  Similarly, when
$\sqrt{2}/3<\delta<2\sqrt{2}/3$, a node only interacts with its
neighborhood.  As the value of $\delta$ grows, nodes can influence
more and more nodes and thus the number of leaders and communities
decreases. Finally, as $\delta\geq\sqrt{D}/3$, $D$ is the diameter
of network, every pair of nodes can influence each other no matter
how far they are.

\begin{figure}
\centering
\subfigure[]{
    \label{fig:subfig:3a} 
    \setcounter{subfigure}{1}
    \includegraphics[width=5cm,height=4.5cm]{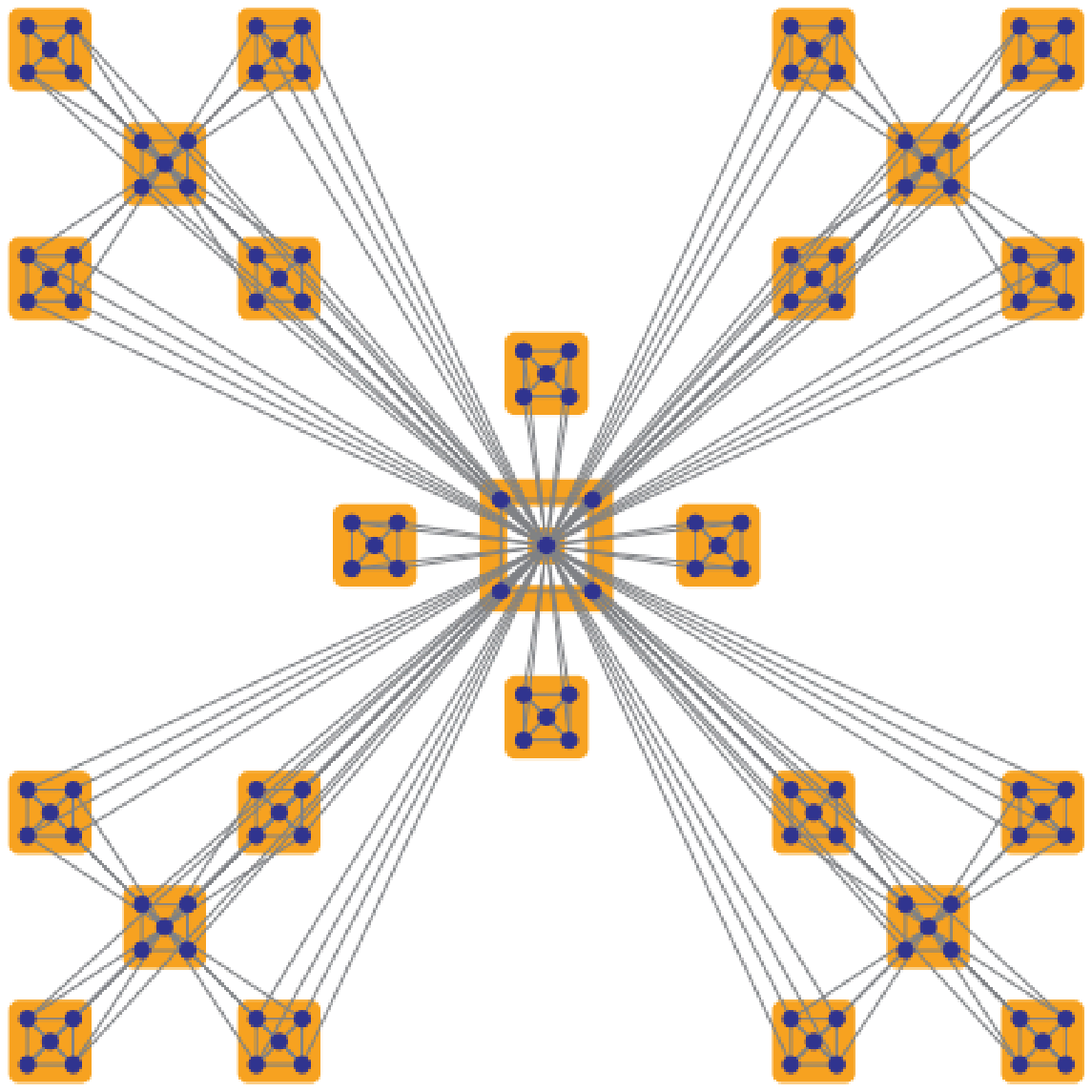}}
  \subfigure[]{
    \label{fig:subfig:3c} 
    \setcounter{subfigure}{2}
    \includegraphics[width=7.5cm,height=5cm]{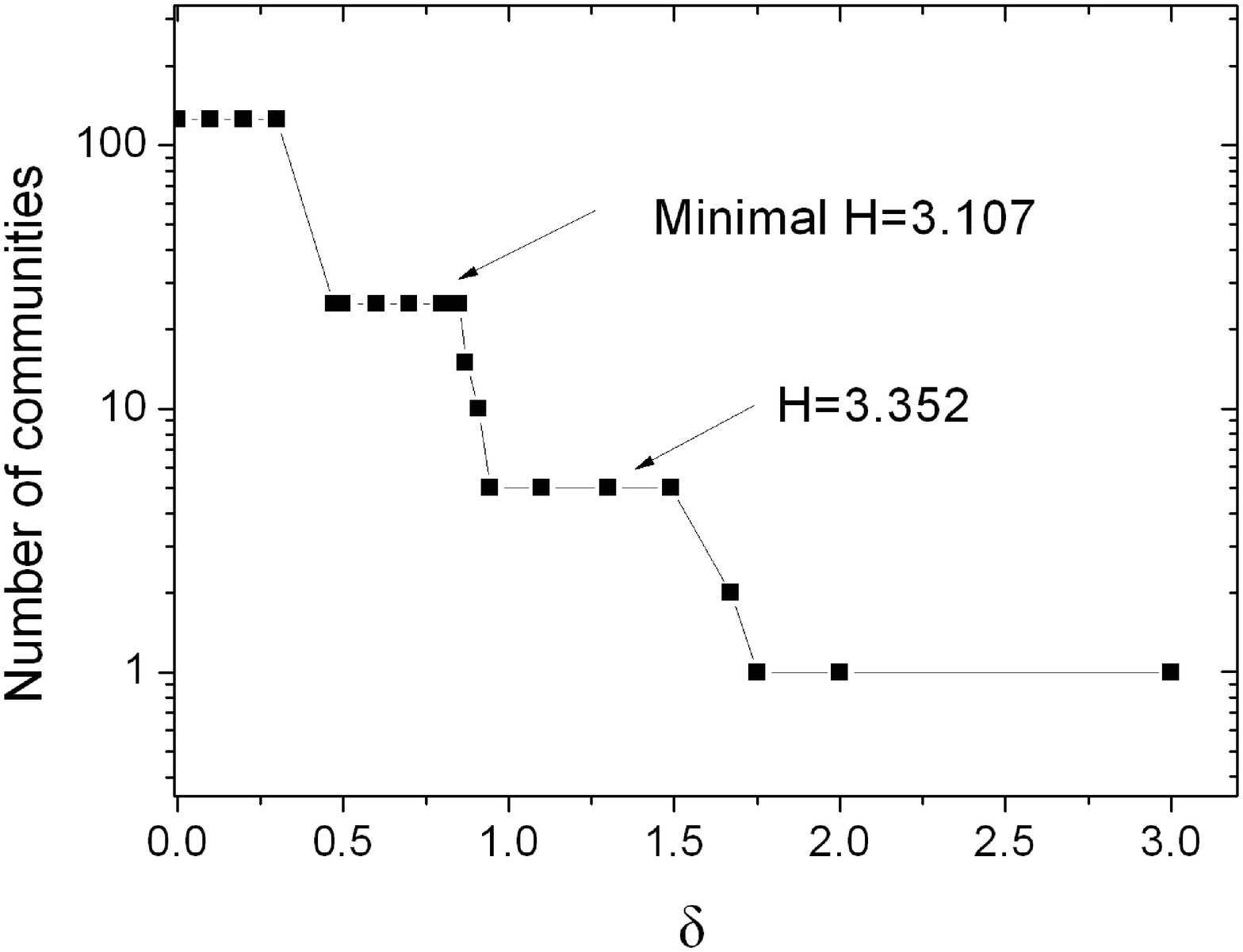}}
\caption{ RB 125 corresponds to the hierarchical scale-free network.
(a) corresponding to 25, 5 modules are the most reasonable
partition in terms of resolution with $H=3.107$ and 3.352. (b) plots
the number of communities versus the value of $\delta$.}
\label{Fig:3}
\end{figure}

To show our method can discover multi-scale community structure with
the variation of $\delta$, we have tested the multi-scale modular
structure in a classical hierarchical scale-free network with 125
nodes, RB125, proposed by Ravasz and Barab\'{a}si\ \cite{Ravasz}. In
Fig.\ref{fig:subfig:3a} we plot the
modular structure found with minimal entropy $H=3.107$ and another
small value $H=3.352$, which shows two different scales that deserve
discussion. The value of $H$ versus different $\delta$ is plotted in
Fig.\ref{fig:subfig:3c}. We observe clearly persistent structures in
25 and 5 communities respectively, that accounts for the
subdivisions more significant in the process, showing two
hierarchical levels for the structure. The partition in 25 modules
and the partition in 5 modules are highlighted on the original
network.

\subsection{ Determining the leaders using local information }

In the algorithm, leadership of one node can be determined through
only local information. By detecting the leader in a community we
gain very useful information of the most influential node in its
community. By removing the leader it can be expected for the
community to suffer serious consequences, like splitting into
several smaller communities. The leader's hierarchy, or the leader's
community, is the area where the leader's opinion is the most
influential opinion. For example, this can be used for an
immunization for epidemic spreading. Thus, the algorithm can naturally determine the number of leaders,
that is also, the number of communities. One interesting feature of the
algorithm is that although it automatically detects the best
leaders, one can manually specify particular nodes as leaders and
build community structures around them.

\subsection{ Determining the overlapping nodes }

It is worthwhile to point out that the vast majority of community
detection methods assume that communities of complex networks are
disjoint, placing each node in only one non-overlapping cluster.
Generally, we call these methods "hard-partition" algorithms.
However, in many real networks communities often overlap to some
extent. An important property of our algorithm is the computation of
a membership vector for each node. Instead of having one number
denoting its membership in a single community, we have a percentage
for each community. As a result, we can easily identify nodes that
naturally belong to more than one community known as overlapping
nodes\cite{Chen}\cite{Shang}\cite{Our1}. So our method is a ``soft-partition"
algorithm. Additionally, we can find nodes that are good followers
of their leader, and also nodes that have no distinguished leader
and serve as a proxy between several communities.

\subsection{ Computational complexity }

The overall complexity of the algorithm depends on the highest
complexity of the three parts of the algorithm. In the following we
analyze each of them sequentially.

The first step is calculating node's leadership $f(i)$. We need to
calculate the exponential function within length of shortest path
$d_{ij}\leq\lfloor\frac{3\delta}{\sqrt{2}}\rfloor$ between pair of
nodes and the complexity of this procedure is at least $O(m)$, $m$
is the number of links. Actually, the computation complexity is
worst, $O(n^2)$, for a dense graph. Next step, determining the
leader nodes of communities, is proceeded by searching all local
highest leadership nodes. This can be done by a simple breadth first
search and the complexity is $O(m)$. The last operation is very
similar to the consensus linear process, whose complexity is $O(n)$
similar to the random walk process.

To conclude this section, the one with the highest computational
complexity is the first step, i.e., calculating the leadership of
nodes. Its complexity depends on the degree of connectivity and the
graph which is very densely connected needs more complexity. This
accounts for the overall complexity of the algorithm is $O(m)$ at
best and $O(n^2)$ at worst.

\section{ Experiments }

In this section, we respectively apply the algorithm to simulated
benchmark networks (LFR networks)\cite{Lancichinetti} and some real social networks: the karate
club network of Zachary\cite{Zachary}, the scientific collaboration network\cite{Girvan} and
finally a large scale semantic network\cite{Palla2}. Results show that the
algorithm can discover multi-scale communities efficiently and
accurately.

\subsection{ The benchmark network }

We empirically demonstrate the effectiveness of the algorithm
through comparison with other five well-known algorithms on the
artificial benchmark networks. These algorithms include:  Newman's
fast algorithm\cite{Newman}, Danon et al's method\cite{Danon}, the
Louvain method\cite{Blondel}, Infomap\cite{Rosvall} and the clique
percolation method\cite{Palla}. We utilize the LFR benchmark
proposed by Lancichinetti et al in\cite{Lancichinetti}. This
benchmark provides networks with scale-free distributions of node
degree and community size and thus poses a much more severe test to
community detection algorithms than standard benchmarks. Many
parameters are used to control the generated networks in this
benchmark: the number of nodes $N$, the average node degree $\langle
k \rangle$, the maximum node degree $max_k$, the mixing ratio $\mu$
(each vertex shares a fraction $\mu$ of its edges with vertices in
other communities), the minimum community size $min_c$ and the
maximum community size $max_c$. The value of $\mu$ varies within
$\lfloor 0,1 \rfloor$ and determines the level of the fuzziness of
the communities in the network. The larger the $\mu$, the more fuzzy
the communities. In our test, we use the default parameter
configuration where $N=1000$, $\langle k \rangle=15$,  $max_k=50$,
$min_c$=20 and $max_c$=50.



\begin{figure}
\centering
\resizebox{0.9\columnwidth}{!}{\includegraphics{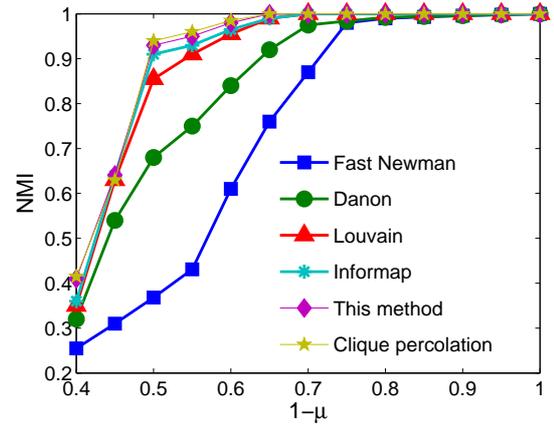} }
\caption{The comparison of NMI with six algorithms.}
\label{Fig:4}       
\end{figure}

To evaluate a community detection algorithm, we use the normalized
mutual information (NMI) measure\cite{Lancichinetti01} to estimate
the partition found by each algorithm. The test focuses on whether
the intrinsic scale can be correctly uncovered. The experimental
results are displayed in Fig.\ref{Fig:4}, where y-axis represents
the value of NMI calculated by the algorithms mentioned above, and
each point in curves is obtained by averaging the values obtained on
50 synthetic networks sampled from above model. As we can see, all
algorithms work very well when $1-\mu$ is more than 0.7 with NMI
larger than $0.85$. Compared with other five algorithms, our
algorithm performs quite well and its accuracy is only slightly
worse than that of the clique percolation in the case of $0.5\leq
1-\mu \leq 0.65$. However, clique percolation is nearly same as the
Breath First Search($BFS$) and very time consuming. The complexity
of clique percolation is almost $O(n^3)$ and much larger than our
method.

As real networks may have some different topological properties from
synthetic ones, in the following we consider several widely used
real-world networks to further evaluate the performance of our
method.

\subsection{ The karate club network of Zachary }

Over the course of two years in the early 1970s, Wayne Zachary
observed social interactions between the members of a karate club at
an American university\cite{Zachary}. He constructed networks of
ties between members of the club based on their social interactions
both within the club and away from it. By chance, a dispute arose
during the course of his study between the club's administrator and
its principal karate teacher over whether to raise club fees, and as
a result the club eventually split into two, forming two smaller
clubs, centered around the administrator and the teacher.

\begin{figure}
\centering
\subfigure[]{
    \label{fig:subfig:5a} 
    \setcounter{subfigure}{1}
    \includegraphics[width=8cm,height=4.5cm]{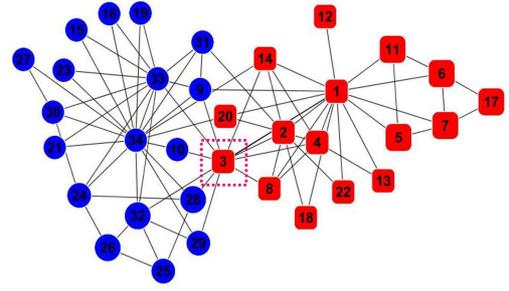}}
  \subfigure[]{
    \label{fig:subfig:5b} 
    \setcounter{subfigure}{2}
    \includegraphics[width=8cm,height=4.5cm]{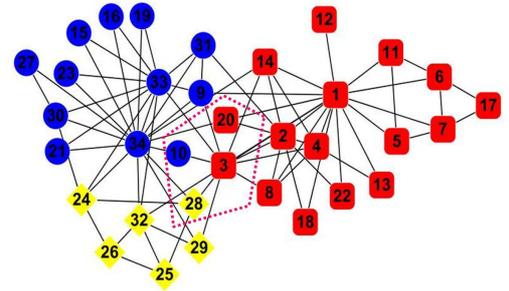}}
  \subfigure[]{
    \label{fig:subfig:5c} 
    \setcounter{subfigure}{3}
    \includegraphics[width=8cm,height=4.5cm]{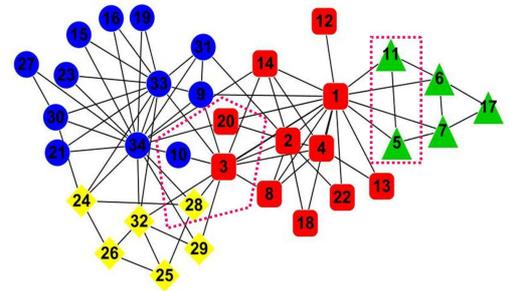}}
\caption{The community structure of the karate club network detected
when $\delta$ (a) equals to the optimal value 1.85, (b) decreases to
1.41 from optimal, and (c) further decreases to 0.933. In subgraphs
(a), (b) and (c), communities are represented by different shapes
and overlapping nodes are enclosed in dashed curves.} \label{Fig:5}
\end{figure}

We minimize the function of $H(\delta)$ and get the optimal value of
$\delta=1.85$ and $H=3.914$.  As it is shown on
Fig.\ref{fig:subfig:5a}, the partition found by our algorithm not
only matches the original partition, but also identifies the exact
leaders. Nodes 1 and 33 own the local highest leadership, which
respectively represent the administrator and the teacher. In this
instance node 3 is detected as an overlapping node because its
membership belonging to two communities is nearly equal. Actually,
node 3 is on the border between the communities and so it is
understandable that it might be an ambiguous case.

Compared with the optimal situation, when decreasing $\delta$ to
$1.41$, the entropy $H=4.139$ is also very small. The community
structure detected in this situation is shown in
Fig.\ref{fig:subfig:5b}, which reveals another scale of
relationships among the members of the karate club. Node 28 becomes
another local highest leadership node and four most unstable nodes
including nodes 3, 10, 20, 28 are marked in a dashed curve. Such
members have good friendship with more than one clubs at the same
time, so they are overlapping nodes in this situation. And now the
number of communities detected in the karate network is three.
Furthermore, as decreasing $\delta$ to $0.93$ $H$ becomes $4.436$,
we get a partition with 4 communities shown in
Fig.\ref{fig:subfig:5c}. This partition is identical to
\cite{Newman04} described by Newman. Six overlapping nodes are
detected which constitute the fuzzy boundaries of the communities.
Thus, partitions using different scales of $\delta$ are able to
reflect multi-scale property of the real networks.

\subsection{ The scientific collaboration network }

The scientific collaboration network was collected by Girvan and
Newman\cite{Girvan} and has been examined in Refs
\cite{Our1}\cite{Our3}. This network consists of 118 nodes
(scientists or authors), and edges between them indicate
co-authorship of one or more papers appearing in the archive. The
collaborative ties represented in the figure are not limited to
papers on topics concerning networks -- we were interested primarily
in whether people know one another, and collaboration on any topic
is a reasonable indicator of acquaintance.

\begin{figure}
\centering
\subfigure[]{
    \label{fig:subfig:6a} 
    \setcounter{subfigure}{1}
    \includegraphics[width=7.5cm,height=5cm]{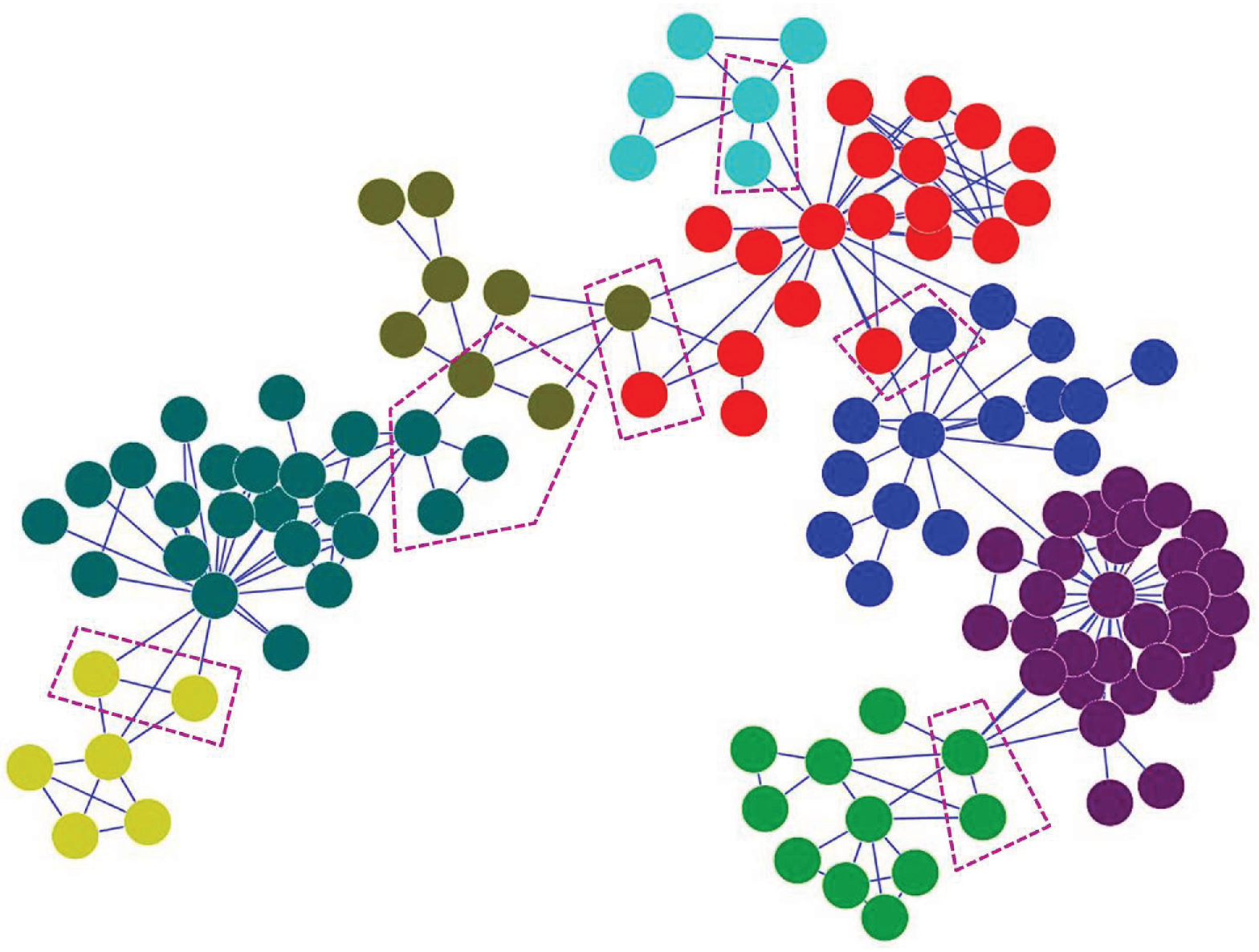}}
  \subfigure[]{
    \label{fig:subfig:6b} 
    \setcounter{subfigure}{2}
    \includegraphics[width=7.5cm,height=5cm]{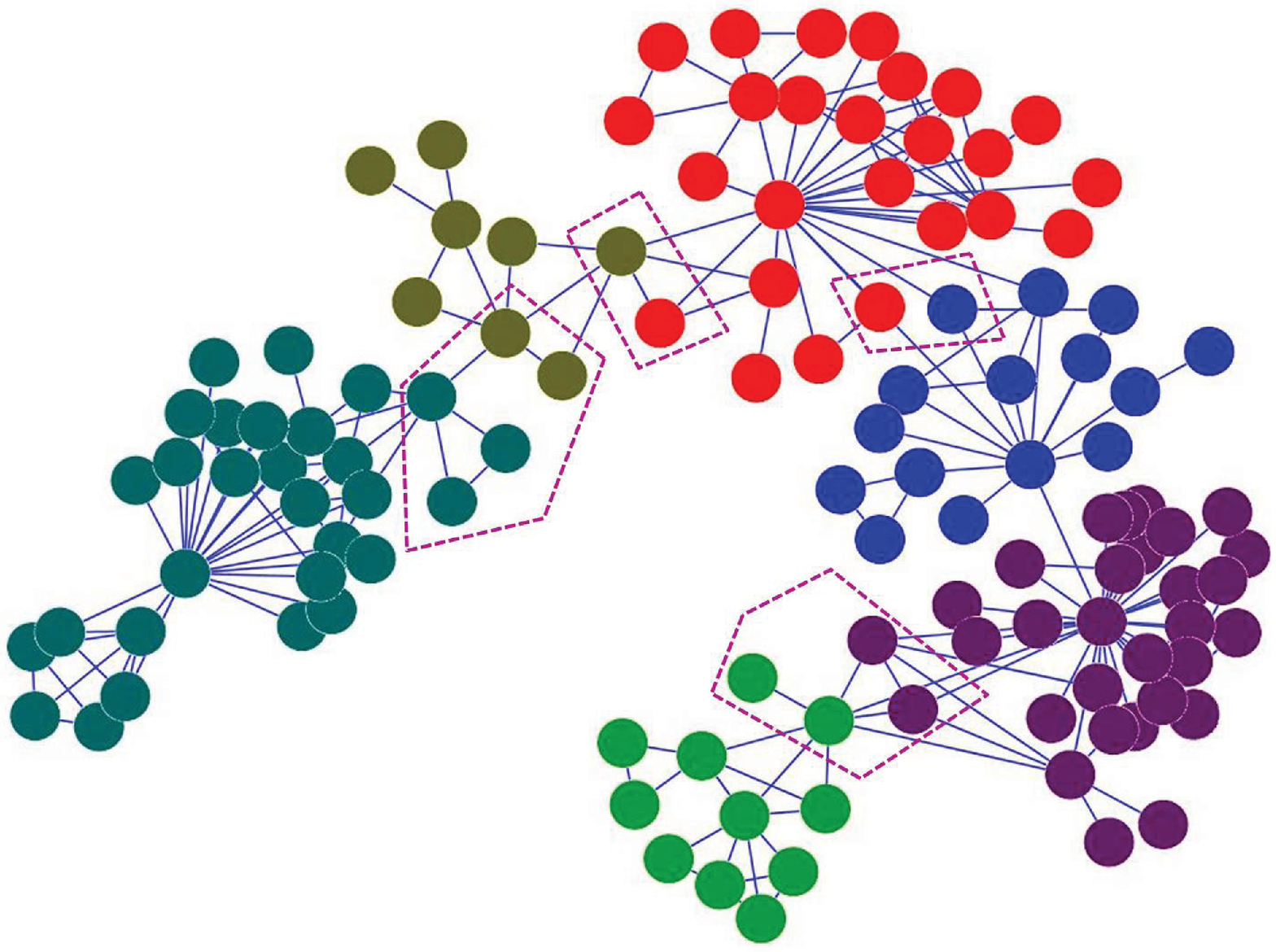}}
\caption{The community structure of the scientific collaboration
network obtained when $\delta$ (a) equals to the optimal value
1.493, (b) is amplified to 1.749 from optimal. In both subgraphs (a)
and (b), overlapping nodes are enclosed in dashed curves.}
\label{Fig:6}
\end{figure}

The present method detects eight communities with optimal
$\delta=1.493$ and minimal entropy $H=5.447$.
Fig.\ref{fig:subfig:6a} shows the community structure detected at
optimal situation which is exactly same as
Refs.\cite{Girvan}\cite{Our1}. This confirms our partition as a good
one. However, we believe our method can also make a meaningful
``coarse-grained" partition which is visually reasonable. So the
value of $\delta$ is amplified to 1.749 from optimal and the
corresponding entropy $H=6.483$. Owing to the the amplification of
the influence range of nodes with $\delta$, the number of
communities decreases. From Fig.\ref{fig:subfig:6b}, we notice some
``uninfluential" communities, like the light blue and yellow ones,
are merged by the more powerful red and dark green communities,
respectively. Finally we get six communities which can be
interpreted readily by the human eye. These multi-scale partitions
will be invaluable in helping us to understand the large-scale
structure of these network data. Furthermore, overlapping nodes
enclosed in dashed curves in Fig.\ref{Fig:6} are detected according
to their membership vectors. These nodes generally locate on the
borders of two or more communities and represent authors with
multiple research interests or cross-discipline background. Maybe
such nodes play a role in bridging two or more communities in a
complex network of other types. The ability to find overlapping
nodes is a distinguished feature of our method and useful to reveal
a natural characteristic in many social networks.

\subsection{ A large scale semantic network }

\begin{figure}
\centering
\resizebox{1\columnwidth}{!}{\includegraphics{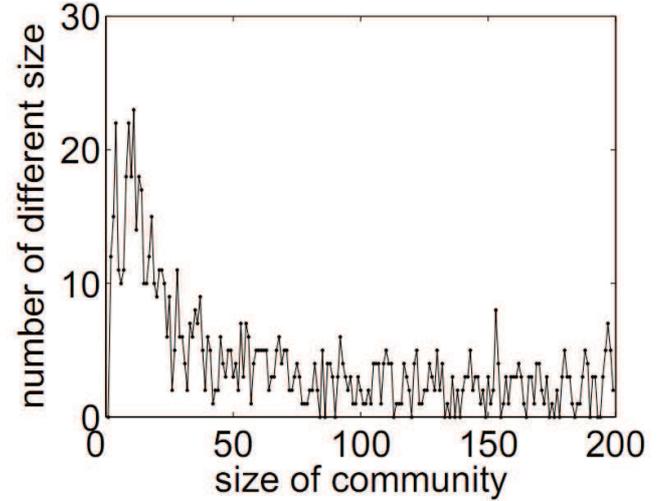} }
\caption{The distribution of community size in a linear plot.}
\label{Fig:7} 
\end{figure}

The semantic network from Ref.\cite{Palla2} contains 7207 phrases
and 31784 edges. The weights of edges are calculated in terms of
phrase co-occurrences. For visualization purpose, our algorithm
outputs a transformed adjacency matrix (in which the vertices within
the same communities have been arranged together) with a
hierarchical community structure. The distribution of community sizes is
shown in Fig.\ref{Fig:7}. Totally, 569 communities are
detected by setting optimal $\delta=2.931$ and minimal entropy
$H=5.952$. The maximum size of community is 139, the minimum size is
2, and the average size is 12.57. One can see an approximate
power-law phenomenon, that is, most communities are small and only a
few are big. Among them, we have selected four interesting
communities listed as follows:

\textbf{Community 1} = \{Scientist, Inventor, Genius, Gifted,
Brilliant, Intelligent, Smart, Science, Intelligence, Musician\};

\textbf{Community 2} = \{Violin, Instrument, Cello, Band, Tuba,
Clarinet, Orchestra, Trumpet, Trombone, Oboe, Woodwind, Symphony,
Flute, Bass, Viola, Fiddle\};

\textbf{Community 3} = \{Ovation, Sitting, Low, Descent, Up, Step,
Ascend, Elevator, Ascent, Staircase, Stairwell, Climb, Steps,
Ladder, Stairs, Wake, Stairway, Rise, Escalator, Stair, Down,
Standing, Resting, Using\};

\textbf{Community 4} = \{Nails, Hammer, Carpenter, Screw,
Screwdriver, Tool, Pliers, Wrench, Sickle, Mechanic, Phillips\}.

These four communities are all reasonable modules listed in
Ref.\cite{Palla2} and the elements of each are all have same
meaning. Among these elements, $\{Musician, Intelligence\}$ are
uncovered as overlapping nodes between communities 1 and 2, and
$\{Using, Tool, Mechanic\}$ are the overlapping nodes between
communities 3 and 4. We can easily recognize that these overlapping
phrases have fuzzy meanings and have high value of phrase
co-occurrences.

\begin{figure}
\centering
\resizebox{1\columnwidth}{!}{\includegraphics{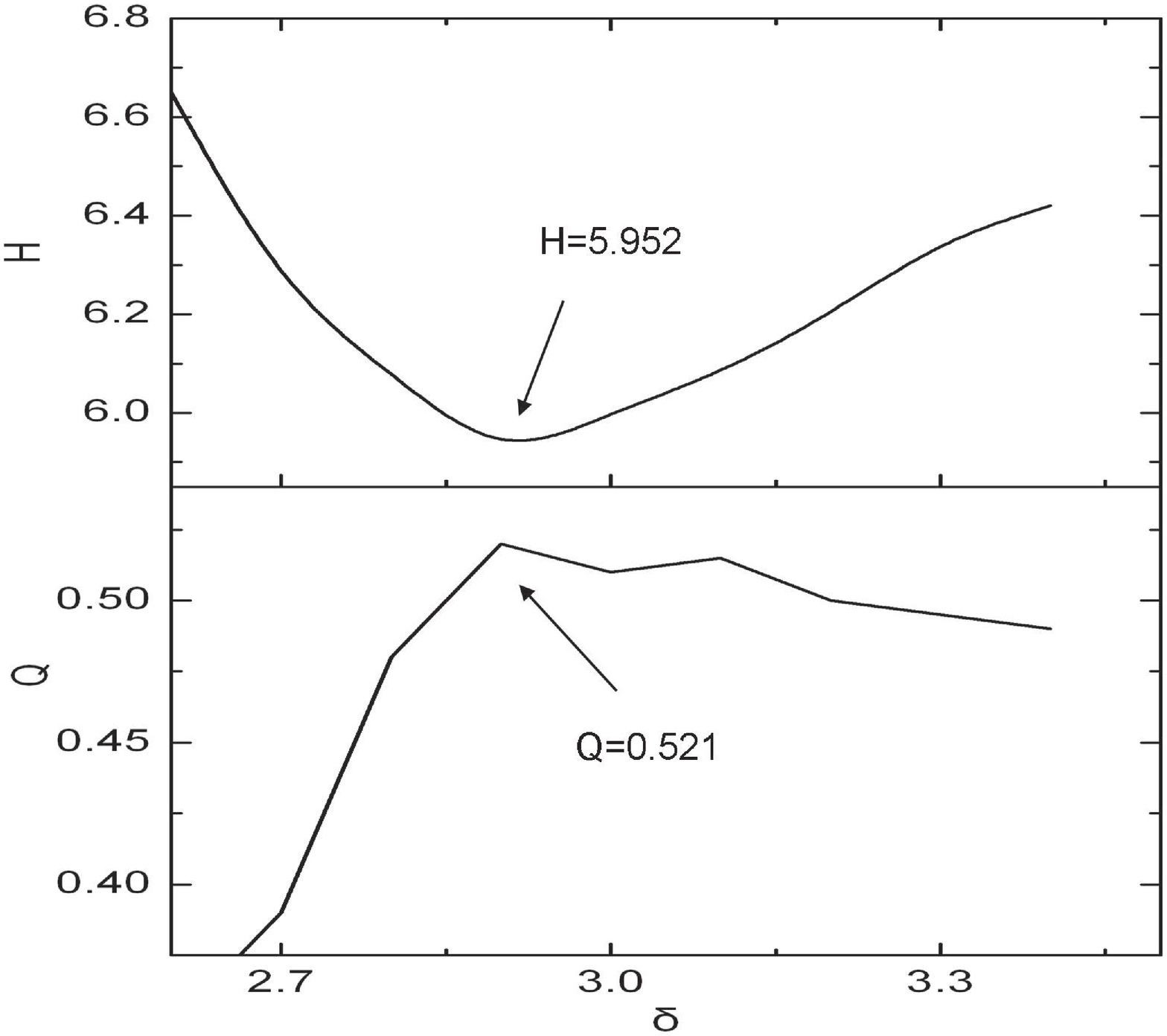} }
\caption{The comparison of Modularity $Q$ with topological entropy $H$ across multi-scale $\delta$.}
\label{Fig:8} 
\end{figure}

As the inherent community structure for this large semantic network
is usually unknown, it is worth to make use of a measure to
quantitatively evaluate the performance of our method. Here the
popular modularity Q\cite{Newman01}\cite{Newman02} is adopted as a
reference, which was proposed by Newman and Girvan and has been
heavily used for community detection in recent years. $Q$ is defined
as:

\begin{equation} \label{eq:4}
Q=\sum_{i=1}^{c}[\frac{l^{in}_i}{L}-(\frac{d_i}{2L})^2],
\end{equation}
Here, $c$ is the number of communities, $L$ is the total number of
edges in the network, and $l^{in}_i$ and $d_i=2 l^{in}_i+l^{out}_i$
are the number of edges and the sum of vertex degrees in the $i$th
community, respectively. Fig.\ref{Fig:8} shows the result that
compares modularity $Q$ with the topological entropy $H$ across
multi-scale of $\delta$. As we can see, the main trend is that the
lower value of $H$, the larger value of $Q$. When $\delta$ reaches
the optimal value $H=5.952$, the Modularity $Q$ also reaches the
maximal $Q=0.521$ exactly. The result shows the community structure
of the network corresponding to a certain $\delta$ is strong and
robust. In conclusion, our algorithm can uncover the most suitable
community scale effectively on real-world networks.

\section{ Conclusion }

In summary, we have presented a novel community detection method
based on local information in social networks. The algorithm does
not embrace the universal approach but tries to focus on local
social ties and model multi-scales of social interactions that occur
on those networks. It identifies leaders and then detects
communities located around the leaders using random walk dynamic.
Our method not only supports overlapping communities detection using
a membership vector to denote node's involvement in each community,
but can also describe different multi-resolution clusters allowing
to discover ``coarse-grained" modules versus the optimal partition.
Applying our algorithm to several typical real-world networks with
well defined community structures, we obtained reasonable results.
So this method is feasible to be used for the accurate detection of
community structures in complex networks. To sum up, from a new
perspective, we propose a new community detection algorithm based on
local information in this paper. The computational results on real
social networks show that the new method not only can detect the
accurate communities but also can extract the hierarchical
structures of the networks.

\vskip 1mm \vspace{0.3cm}

\noindent{\bf Acknowledgments:} We are grateful to the anonymous reviewers
for their valuable suggestions which are very helpful for improving the manuscript.
The authors are separately supported by NSFC grants 11131009, 60970091, 61171007,
91029301, 61072149, 31100949, 61134013 and grants kjcx-yw-s7 and KSCX2-EW-R-01 from CAS.
This research is also partially supported by Shanghai Pujiang Program and
the Aihara Project of the FIRST program from JSPS initiated by CSTP.\\


\begin{thebibliography}{1}

\bibitem{Barabasi} Barab\'{a}si.A.L, Albert.R, {\em Science}, \textbf{286}, (1999) 509-512.

\bibitem{Albert} Albert.R, Barab\'{a}si.A.L, Jeong.H, {\em Nature }, \textbf{401}, (1999) 130.

\bibitem{Li} Li.X.G., Gao.Z.Y, Li.K.P, Zhao.X.M, {\em Phys. Rev. E}, \textbf{76}, (2007) 016110.

\bibitem{Liljeros} Liljeros.F, Edling.C.R, Amaral.L.A.N, Stanley.H.E, Aberg.Y, {\em Nature }, \textbf{411}, (2001) 907-908.

\bibitem{Sumiyoshi} Sumiyoshi.A, Norikazu.S, {\em Phys. Rev. E}, \textbf{74}, (2006) 026113.

\bibitem{Newman} Newman.M.E.J, {\em Phys. Rev. E}, \textbf{69}, (2004) 066133.

\bibitem{Newman01} Newman.M.E.J, Girvan.M, {\em Phys. Rev. E}, \textbf{69}, (2004) 026113.

\bibitem{Newman02} Newman.M.E.J, {\em Proc. Natl. Acad. Sci}, \textbf{103}, (2006) 8577-8582.

\bibitem{Danon} Danon.L, Duch.J, Guilera.D, Arenas.A, {\em J. Stat. Mech}, \textbf{29}, (2005) P09008.

\bibitem{Clauset} Clauset.A, Newman.M.E.J, Moore.C, {\em Phys. Rev. E}, \textbf{70}, (2004) 066111.

\bibitem{Newman03} Newman.M.E.J, {\em Phys. Rev. E}, \textbf{74}, (2006) 036104.

\bibitem{Newman04} Newman.M.E.J, {\em Eur.Phys.J.B}, \textbf{38}, (2004) 321-330.

\bibitem{Ravasz} Ravasz.E, Barab\'{a}si.A.L, {\em Phys. Rev. E}, \textbf{67}, (2003) 026112.

\bibitem{Lambiotte} Lambiotte.R, Delvenne.J.C, Barahona.M, (2009), arXiv:0812.1770.

\bibitem{Baras} Baras.J.S, Hovareshti.P, (2008), {\em Proceedings of 47th IEEE Conference on Decision and Control}, 2973-2978.

\bibitem{Mucha} Mucha.P.J , Richardson.T, Macon.K, Porter.M.A, Jukka-Pekka Onnela, {\em Science}, \textbf{328}, (2010) 876-878.

\bibitem{Palla} Palla.G, Der\'enyi.I, Farkas.I, Vicsek.T, {\em Nature}, \textbf{435}, (2005) 814-818.

\bibitem{Our1} Li.H.J, Wang.Y, Wu.L.Y, Liu.Z.P, Chen.L, Zhang.X.S, {\em Europhysics Letters }, \textbf{97}, (2012) 48005.

\bibitem{Our} Zhang.X.S, Wang.R.S, Wang.Y, Wang.J, Qiu.Y, Wang.L, Chen.L, {\em Europhysics Letters}, \textbf{87}, (2009) 38002.

\bibitem{Zachary} Zachary.W.W, {\em Journal of Anthropological Research}, \textbf{33}, (1977) 452-473.

\bibitem{Palla2} Palla.G, Barab\'{a}si.A.L, Vicsek.T, {\em Nature}, \textbf{446}, (2007) 664-667.

\bibitem{Girvan} Girvan.M, Newman.M.E.J, {\em Proc.Natl.Acad.Sci}, \textbf{99}, (2002) 7821-7826.

\bibitem{Our2} Li.Z.P, Zhang.S.H, Wang.R.S, Zhang.X.S, Chen.L, {\em Phys. Rev. E}, \textbf{77}, (2007) 036109.

\bibitem{Gfeller} Gfeller.D, Chappelier.J.C, Los Rios.P.De, {\em Phys.Rev.E} \textbf{72(5)}, (2005) 056135.

\bibitem{Bianconi} Bianconi.G, Pin.P, Marsili.M, {\em Proc.Natl.Acad.Sci}, \textbf{106(28)}, (2009) 11433-11438.

\bibitem{Chen} Chen.D.B, Shang.M.S, Fu Y, {\em Physica A}, \textbf{389}, (2010) 4177-4187.

\bibitem{Shang} Shang.M.S, Chen.D.B, Zhou.T, {\em Chin.Phys.Lett}, \textbf{27}, (2010) 058901.

\bibitem{Blondel} Blondel.V.D, Guillaume.J.L, Lambiotte.R, Lefebvre.E, {\em J. Stat. Mech}, \textbf{10}, (2005) P10008.

\bibitem{Rosvall} Rosvall.M, Bergstrom.C.T, {\em Proc.Natl.Acad.Sci}, \textbf{105(4)}, (2008) 1118-1123.

\bibitem{Lancichinetti} Lancichinetti.A, Fortunato.S, {\em Phys. Rev. E}, \textbf{80}, (2009) 016118.

\bibitem{Lancichinetti01} Lancichinetti.A, Fortunato.S, {\em Phys. Rev. E}, \textbf{80}, (2009) 056117.

\bibitem{Our3} Zhang.J.H, Zhang.S.H, Zhang.X.S, {\em Physica A}, \textbf{387}, (2008) 1675-1682.

\end {thebibliography}
\end{document}